\documentclass[12pt]{article}

\title{Method of Hydrodynamic Images and Quantum Calculus  in Fock-Bargmann Representation of Quantum States}

\author{Oktay K Pashaev \\
Department of Mathematics \\ Izmir Institute of Technology \\ Izmir 35430, T\"urkiye}

\begin{document}

\maketitle              

\begin{abstract}
  We propose a new approach to
quantum states in Fock space in terms of classical hydrodynamics. By conformal mapping of complex analytic function, representing the wave function of quantum states in Fock-Bargmann representation, we define the complex potential, describing these quantum states by incompressible and irrotational classical hydrodynamic flow. In our approach, zeros of the wave function appear as a set of point vortices (sources) in plane with the same strength, allowing interpretation of them as images in a bounded domain.  For the cat states we find fluid representation as descriptive of a point source in the oblique strip domain, with infinite number of periodically distributed images. For the annular domain, the infinite set of images is described by Jackson $q$-exponential functions. We show that these functions represent the wave functions of quantum coherent states of the $q$-deformed quantum oscillator in $q$-Fock-Bargmann representation and describe the infinite set of point vortices, distributed in geometric progression.
\end{abstract}

Keywords:{quantum calculus, Fock-Bargmann representation, vortex images, coherent states}


 %
\section{Introduction}
The hydrodynamical analogy is a fruitful approach in many fields of physical science and mathematics, including electromagnetic theory, quantum mechanics, nuclear physics, dynamical systems and complex analysis.
In quantum mechanics, the  hydrodynamic representation of wave function proposed by Madelung \cite{Madelung}, gives intuitive picture of probability flow propagating in quantum fluid. The argument of wave function is identified with the velocity potential of the fluid and every eigenfunction of the Schr\"odinger equation can be interpreted as a special type of stationary flow. By expressing the velocity of fluid as an analytic function and applying the residue theorem, WKB got semiclassical quantization rule, which in fact is known in complex analysis as the argument principle, counting number of zeros and poles. The hydrodynamic interpretation gives meaning to an analytic function as descriptive of incompressible and irrotational fluid in plane and classification of singularities of the function by point vortices, sinks and sources, etc. 

Conformal mappings and method of images are important technical tools to solve boundary value problems in bounded domains.
The method of images was developed by W.Thomson as powerful method of solving electrostatic problems and it was applied to many problems in Maxwell's Treatise on Electricity and Magnetism \cite{Maxwell}.
For simple planar domains like the circle, the wedge, the strip and the annular domain \cite{Poincare}, the distribution of the hydrodynamic images is determined by the circle theorem \cite{Milne}, the wedge theorem \cite{P2}, the strip theorem \cite{P3} and the two circle theorem \cite{P1}, respectively. For the strip domain, the infinite set of periodic images is described by elliptic functions. For the wedge and the annular domain  it can be represented naturally by the $q$-periodic functions of the quantum or the $q$-calculus \cite{Kac}. For real $q$,  the stretching from the origin results from inversion in two circles \cite{Ford}, while for  $q$ as a root of unity, $q^{2n} =1$, the rotation corresponds to reflection in two lines. This is why the geometric inversion in complex plain becomes the
origin of the method of images.
For the wedge domain the kaleidoscope of finite number of images is located at vertices of regular polygon \cite{PashaevKocak}, \cite{KocakPashaev}, and possess the quantum group symmetry \cite{P4}.

In this paper we propose a new approach to
quantum states in Fock space in terms of classical hydrodynamics. By conformal mapping of complex analytic function, representing the a wave function of quantum states in Fock-Bargmann representation, we define the complex potential, describing these quantum states by incompressible and irrotational classical hydrodynamic flow. In our approach, zeros of the wave function appear as a set of point vortices in plane with the same strength. This allows us in some cases to interpret these vortices as images of the one real vortex in a bounded domain, with vanishing normal velocity across the boundary curve.  For the cat states we find fluid representation as descriptive of point source in the oblique strip domain, with infinite number of periodically distributed images.
For the annular domain with real $q$, the infinite set of images is described by Jackson $q$-exponential functions. We show that these functions represent the wave functions of quantum coherent states of the $q$-deformed quantum oscillator in $q$-Fock-Bargmann representation and describe the infinite set of point vortices, distributed along a half-line in geometric progression.

\section{Holomorphic Wave Functions and Hydrodynamic Flow}
The the wave function in quantum mechanics has appeared in Schr\"odinger first communication on wave mechanics \cite{Schrodinger}. By the Log transformation 
$$S = \hbar \ln \Psi$$
in the Hamilton-Jacobi equation with action variable $S$ he introduced
unknown function $\Psi$ , the logarithmic derivative of which then becomes complex velocity vector of Madelung fluid \cite{Madelung}. This suggests us to introduce the following hydrodynamic representation for holomorphic wave functions.

\subsubsection{The Vortex Flow}

{\bf Definition 1.}
The analytic functions
\begin{equation}
f(z) = \frac{i \Gamma}{2\pi} \ln \Psi(z), \hskip1cm \bar{v}(z) = \frac{d f}{dz} = \frac{i \Gamma}{2\pi} \frac{\Psi'(z)}{\Psi(z)}, \label{vortexpotential}
\end{equation}
where $z = x+iy$, represent the vortex type complex potential $f(z)$ of two dimensional incompressible and irrotational fluid flow and corresponding complex velocity $\bar v(z)$.

The function $f$ determines a conformal mapping from $z$-plane to $f(z) = \varphi + i\psi$ plane, and expression for velocity $\bar v(z)$ is holomorphic version of the Cole-Hopf transformation
from the heat equation to nonlinear Burgers equation \cite{PashaevGurkan}.
The holomorphic wave function in terms of complex potential $f$, for the vortex representation (\ref{vortexpotential}) is

\begin{equation}
\Psi(z) = |\Psi(z)| e^{i \arg \Psi} = \exp\left(-i \frac{2\pi}{\Gamma} f(z)\right) = \exp\left(\frac{2\pi}{\Gamma} \psi \right) \exp\left(-i\frac{2\pi}{\Gamma} \varphi \right).
\end{equation}

In this representation,  the velocity potential curves $\varphi = constant$  are orthogonal to the velocity field and preserve the constant argument of $\Psi(z)$.
From another side, the curves $\psi = constant$, along which the modulus of the wave function $\Psi$ is constant, are lines of the stream. Such interpretation of the hydrodynamic flow in terms of the modulus and argument of some complex wave function $\Psi(z)$ was indicated in Poincare lectures on vortex theory in 1893 \cite{Poincare}, much early before discovery of quantum mechanics and the wave function.

{\bf Proposition 1.} 
Every zero $z = z_0$ of the wave function (\ref{vortexpotential}), $\Psi(z_0) =0$, describes the point vortex
with complex potential $f$ and corresponding complex velocity $\bar v$,
\begin{equation}
f(z) = \frac{i \Gamma}{2\pi} \ln (z-z_0),\hskip1cm \bar{v}(z) =  \frac{i \Gamma}{2 \pi } \frac{1}{z - z_0},
\end{equation}
where $\Gamma$ characterises the strength (intensity) of the vortex.

The wave function $w = w(z) = \Psi(z)$ gives conformal mapping from complex plane $z \in C$ to complex plane $w = u + iv \in C$. Then, (\ref{vortexpotential}) represents the point vortex
at the origin in $w$ plane,
$$ f(z(w)) = \frac{i\Gamma}{2 \pi} \ln w.
$$ 
The stream function of this flow
\begin{equation}
\psi(u,v) = \frac{\Gamma}{2\pi} \ln |w| =  \frac{\Gamma}{2\pi} \ln |\Psi(z)|,
\end{equation}
describes the stream lines $\psi(u,v) = c_0$ as circles of radius

\begin{equation}
|\Psi(z)| = e^{\frac{2\pi}{\Gamma} c_0}.
\end{equation}

For the flow in a domain, bounded by curve C, the boundary condition on normal velocity to the curve is zero, which implies that the stream function $\psi = 0$,
and as follows, the wave function $\Psi(z)$ on boundary curve is unimodular, 
$$
|\Psi(z)| =1.$$

\subsubsection{The Source/Sink Flow}

{\bf Definition 2.}
The  analytic functions
\begin{equation}
f(z) =  \frac{N}{2\pi} \ln \Psi(z), \hskip1cm    \bar{v}(z) = \frac{d f}{dz} = \frac{N}{2\pi} \frac{\Psi'(z)}{\Psi(z)},  \label{sourcepotential}
\end{equation}
represent the source/sink type complex potential $f(z)$ of the two dimensional flow and corresponding complex velocity $\bar v(z)$.

The wave function for the source/sink field representation is
\begin{equation}
\Psi(z) = \exp{\frac{2\pi}{N} f} = \exp \left(\frac{2\pi}{N} \varphi(x,y)\right) \exp \left(i \frac{2\pi}{N} \psi(x,y)\right).
\end{equation}
The function gives conformal mapping to complex plane $w = \Psi(z)$, so that in this plane (\ref{sourcepotential}) represents the point source/sink at the origin
$$f(z(w)) =  \frac{N}{2\pi} \ln w. $$
The stream lines of this flow in $w$ plane correspond to constant  argument of the wave function
$$
 \frac{N}{2\pi} \arg \Psi(z) = \psi(u,v) = c_0. $$

{\bf Proposition 2.}
Every zero $z = z_0$ of the wave function (\ref{sourcepotential}), $\Psi(z_0) =0$, corresponds to the point source/sink
with complex potential $f$ and complex velocity $\bar v$,
\begin{equation}
f(z) = \frac{N}{2\pi} \ln (z-z_0),\hskip1cm \bar{v}(z) =  \frac{N}{2 \pi } \frac{1}{z - z_0},
\end{equation}
where $N$ characterizes the strength (abundance) of the source ($N > 0$) or the sink ($N < 0$).

\subsubsection{The Source and Vortex Flow}

The analytic function
$$
f(z) =  \frac{N + i\Gamma}{2\pi} \ln \Psi(z)
$$
represents complex potential of a source and a vortex flow. The complex velocity for this flow is
$$
\bar v(z) = \frac{df}{dz} = \frac{N + i\Gamma}{2\pi} \frac{\Psi'(z)}{\Psi(z)}.
$$

\subsection{Boundary Conditions}
For a domain bounded by curve $C$ the normal velocity of the flow vanishes at the boundary. This implies the boundary condition
as the reality condition for complex potential
$$
\Im f(z)|_C = \psi|_C =0.
$$
For the vortex flow (\ref{vortexpotential}) on $C$, this gives
$$
\psi(x,y) = \frac{\Gamma}{4\pi} \ln |\Psi(z)| =0
$$
or the unimodularity condition on the wave function
$$
|\Psi(z)|^2 =1,
$$
which describes the unit circle in $w$ plane.
For the source/sink flow (\ref{sourcepotential}) on $C$,
$$
\psi(x,y) = \frac{N}{4\pi i} \ln \frac{\Psi(z)}{\overline{\Psi(z)}} =0,
$$
this gives the reality condition on the wave function
$$
\Psi(z) = \overline{\Psi(z)}.
$$
As a result we have following proposition.

{\bf Proposition 3.}
The boundary condition for the two dimensional flow, as vanishing of normal velocity $v_n$ on the boundary curve $C$,
corresponds to boundary condition on the wave function:

1) for the vortex flow

\begin{equation}
|\Psi(z)|^2 =1,
\end{equation}

2) for the source/sink flow

\begin{equation}
\Psi(z) = \overline{\Psi(z)}.
\end{equation}

\subsection{Normalization Freedom}
The complex potential $f(z)$ of the flow with complex velocity $\bar v(z) = df/dz$ is determined up to an arbitrary complex constant:
$f(z) \rightarrow f(z) + c_0$. This implies that the wave function corresponding to the given flow $\bar v (z)$ is defined up to an arbitrary
(frequently an infinite) multiplication constant $B$: $\Psi(z) \rightarrow B \,\Psi(z)$.

\section{The Wedge Theorem and Wave Function}
The wedge theorem \cite{P4} describes the flow in the wedge domain.

{\bf Theorem 1.}
For a given in plane flow $f(z)$, introduction of the boundary wedge domain with angle $\alpha = \frac{\pi}{n}$, produces the flow
\begin{equation}
F_q (z) = \sum^{n-1}_{k=0} f(q^{2k}z) + \sum^{n-1}_{k=0} \bar{f}(q^{2k}z).
\end{equation}
where $q = e^{i \frac{\pi}{n}}$ is primitive root of unity, so that $q^{2n} =1$. The flow is $q^2$-periodic (rotation invariant):
$$ F_q(q^2 z) = F_q(z). $$

This theorem implies following representation for the wave function in the wedge domain.

{\bf Theorem 2.}
The wave function $\Psi_q$ in the wedge domain with angle $\alpha = \frac{\pi}{n}$ and $q = e^{i \frac{\pi}{n}}$, in vortex representation (\ref{vortexpotential}) is
\begin{equation}
\Psi_q(z) = \prod^{n-1}_{k=0} \frac{\Psi(q^{2k} z)}{\overline{\Psi}(q^{2k} z)}, \label{vortexfunction}
\end{equation}
and in source/sink representation (\ref{sourcepotential})  it is
\begin{equation}
\Psi_q(z) = \prod^{n-1}_{k=0} \Psi(q^{2k} z) \,\overline{\Psi}(q^{2k} z), \label{sourcefunction}
\end{equation}
where function $\Psi(z)$ corresponds to the whole complex plane. These functions are $q^2$-periodic:
$$
\Psi_q(q^2 z) = \Psi_q(z).
$$

{\bf Proof}:
The wave function (\ref{vortexfunction}) on boundaries of the wedge domain $z =x$ and $z = q x$, satisfy following equations, respectively,
$$ \overline{\Psi_q(x)}   = \frac{1}{\Psi_q(x)},   \hskip1cm   \overline{\Psi_q(q x)}   = \frac{1}{\Psi_q(qx)},                     $$
which means that the function on the wedge boundaries is the unimodular
$$|\Psi_q(x)|^2 =1, \hskip1cm |\Psi_q(qx)|^2 =1.
$$
The wave function (\ref{sourcefunction}) on boundaries of the wedge domain $z =x$ and $z = q x$, is equal, respectively,
\begin{equation}
\Psi_q(x) = \prod^{n-1}_{k=0} |\Psi(q^{2k} x)|^2, \hskip1cm \Psi_q(q x) = \prod^{n-1}_{k=0} |\Psi(q^{2k+1} x)|^2.
\end{equation}
It is the real and non-negative valued function on the boundaries.

\section{The Strip Theorem and Wave Function}
For the hydrodynamic flow in a strip domain we have the strip theorem \cite{P3}.

{\bf Theorem 3.}
The complex potential for incompressible and irrotational flow in the strip domain $S:\{ z = x+i y;   -h/2 \le y \le h/2          \}     $ is
\begin{equation}
F(z) = \sum^\infty_{n=-\infty} f(z + (2n) i h) + \sum^\infty_{n=-\infty} \bar{f}(z + (2n-1) i h),
\end{equation}
where $f(z)$ is potential of the flow in whole plane. It satisfies the boundary conditions: $\Im F(z)|_{z = x \pm i\frac{h}{2}} =0$,
periodicity and the combined periodicity conditions, respectively:
$$
F(z + 2 i h) = F(z), \,\bar F(z + i h) = F(z).
$$

This theorem implies the form of wave functions in the strip domain according to following theorem.

{\bf Theorem 4.}
The wave function corresponding to the strip domain $S$ in vortex representation (\ref{vortexpotential}) is
\begin{equation}
\Psi_h(z) = \prod^{\infty}_{n =-\infty} \frac{\Psi( z + (2 n) i h)}{\overline{\Psi}( z + (2 n -1) i h)}, \label{stripvortexfunction}
\end{equation}
and in source/sink representation (\ref{sourcepotential})  it is
\begin{equation}
\Psi_h(z) = \prod^{\infty}_{n =-\infty} \Psi(z + (2 n) i h) \,\overline{\Psi}(z + (2n -1) i h), \label{stripsourthfunction}
\end{equation}
where function $\Psi(z)$ corresponds to the whole complex plane. The function (\ref{stripvortexfunction}) is inverse combined periodic and the one in (\ref{stripsourthfunction}) is combined periodic, respectively,
\begin{equation}
\bar\Psi_h(z + ih) = \frac{1}{\Psi_h(z)},\hskip1cm \bar\Psi_h(z + ih) = \Psi_h(z).
\end{equation}
Both functions are periodic:
$$
\Psi_h(z + 2 i h) = \Psi_h(z).
$$

{\bf Example 1.}
For the point vortex at the origin in the strip domain $S$,
\begin{equation}
\Psi_h(z) = \prod^{\infty}_{n =-\infty} \frac{( z + (2 n) i h)}{( z + (2 n -1) i h)} = \tanh \frac{\pi}{2h} z.
\end{equation}

{\bf Example 2.}
For the point source at the origin in $S$,
$$
\Psi_h(z) = \prod^{\infty}_{n =-\infty} (z + (2 n) i h) (z + (2n -1) i h) = \prod^{\infty}_{n =-\infty} (z +  n i h) = z \prod^{\infty}_{n =1} (z^2 + n^2 h^2)
$$
it gives (here we neglect irrelevant multiplicative constant)
\begin{equation}
\Psi_h(z) = \sinh \frac{\pi}{h} z. \label{pointsourcestrip}
\end{equation}
As we show in next section, this wave function describes the coherent cat state (\ref{cat1}) in Fock-Barmann representation (\ref{catFB1}), where complex parameter of the coherent state is fixed as
$\alpha = \frac{\pi}{h} $.

\subsection{Oblique Strip Theorem}
For upcoming applications we need a more general version of the strip theorem, with the oblique strip domain.

{\bf Theorem 5.}
The complex potential
\begin{equation}
F_\beta(z) = \sum^\infty_{-\infty} f(z + (2n) i h e^{i\beta}) + \sum^\infty_{-\infty} \bar{f}(z + (2n-1) i h e^{i\beta}), \label{obstriptheorem}
\end{equation}
for the flow in oblique strip domain:
$$S_{\beta}: \{  z = x + i y;   x \tan \beta - \frac{h}{2 \cos \beta}   \le  y \le  x \tan \beta + \frac{h}{2 \cos \beta}\},  $$
between two oblique lines $(-\infty < x < \infty),$
$$ z = x + i ( x \tan \beta - \frac{h}{2 \cos \beta}), and \,\,\,z = x + i ( x \tan \beta + \frac{h}{2 \cos \beta}),$$
is real on these lines and satisfies periodicity and combined periodicity conditions
\begin{equation}
F_\beta(z + 2 i h e^{i\beta}) = F_\beta(z),\hskip1cm \bar F(z + i h e^{i\beta}) = F(z).
\end{equation}

\section{Fock-Bargmann Representation}
Now we are going to  construct the hydrodynamic flow for several types of quantum coherent states.
The pair of bosonic operators $a$ and $a^+$ satisfy the commutation relation
$$
[a, a^+] =1,
$$
and determines the Fock states
\begin{equation}
|n \rangle = \frac{(a^+)^n}{\sqrt{n!}} |0 \rangle, \hskip1cm a |0\rangle =0.
\end{equation}

{\bf Definition 3.}
The Glauber coherent states (not normalized) are vectors in Fock space
 \begin{equation}
|\alpha \rangle = \sum^\infty_{n=0} \frac{\alpha^n}{\sqrt{n!}} |n\rangle, \label{Glaubercoherent}
\end{equation}
where $\alpha \in C$ is an arbitrary complex number.

{\bf Proposition 4.}
The inner product of two coherent states is
$$
\langle \beta | \alpha \rangle = e^{\bar \beta \alpha},
$$
and normalization factor is
$$
\langle \alpha | \alpha \rangle = e^{|\alpha|^2}.
$$

A representation of quantum states in Fock space by complex analytic functions is called the Fock-Bargmann representation \cite{Fock}, \cite{Bargmann}.
Let
$$
|\Psi\rangle = \sum^\infty_{n=0} c_n |n \rangle \label{vectorFock}
$$
is an arbitrary vector in Fock space with normalization condition
$$
\langle \Psi | \Psi \rangle = \sum^\infty_{n=0} |c_n|^2 =1
$$.

{\bf Definition 4.}
The wave function
\begin{equation}
\langle z | \Psi \rangle = \Psi (\bar z),
\end{equation}
where $|z \rangle$ is the Glauber coherent state (\ref{Glaubercoherent}),
determines the entire complex analytic function
\begin{equation}
\Psi(z) = \sum^\infty_{n=0} c_n \frac{z^n}{\sqrt{n!}}.
\end{equation}

{\bf Proposition 5.}
The wave function of the Fock state $|n \rangle$ in Fock-Bargmann representation is monomial
\begin{equation}
\Psi(z) = \frac{z^n}{\sqrt{n!}}.
\end{equation}
It represents the point vortex and point source/sink flow at origin, with strength $n \Gamma$ and $n N$, correspondingly,
\begin{equation}
f_v(z) = \frac{i n \Gamma}{2\pi} \ln z,\hskip1cm f_s(z) = \frac{n N}{2\pi} \ln z.
\end{equation}

{\bf Proposition 6.}
The wave function of the Glauber coherent state $|\alpha \rangle$ in Fock-Bargmann representation is entire function
\begin{equation}
\Psi(z) = e^{\alpha z}.
\end{equation}
It describes homogeneous flow in plane with linear complex potential and constant velocity, determined by complex number $\alpha$,
\begin{eqnarray}
f(z)_{vortex} = \frac{i\Gamma}{2\pi} \alpha\, z, \hskip1cm \bar v(z) =  \frac{i\Gamma}{2\pi} \alpha, \\
f(z)_{source/sink} = \frac{N}{2\pi} \alpha\, z, \hskip1cm \bar v(z) =  \frac{N}{2\pi} \alpha.
\end{eqnarray}

\subsubsection{Displaced Coherent State}
The displaced coherent states are defined as
\begin{equation}
|n; \alpha \rangle = e^{\alpha a^+} |n \rangle,
\end{equation}
and in Fock-Bargman representation their representatives are complex wave functions with translated origin
\begin{equation}
\Psi_n(z; \alpha) = \frac{(z - \bar\alpha)^n}{\sqrt{n!}} e^{\alpha z}.
\end{equation}
The corresponding homogeneous vortex flow
\begin{equation}
f(z) = \frac{i\Gamma n}{2\pi} \ln (z - \bar\alpha) + \frac{i\Gamma}{2\pi} \alpha z
\end{equation}
includes also point vortex with strength $\Gamma n$ at position $z = \bar\alpha$.

\subsection{Cat States Flow}
The even and odd superpositions of two Glauber coherent states
\begin{eqnarray}
| 0 \rangle_\alpha &=& \frac{\cosh \alpha \,a^+}{\sqrt{\cosh |\alpha|^2}} |0\rangle,  \label{cat0}\\
| 1 \rangle_\alpha &=& \frac{\sinh \alpha \,a^+}{\sqrt{\sinh |\alpha|^2}} |0\rangle, \label{cat1}
\end{eqnarray}
are called the cat states. An arbitrary state $|\Psi \rangle$ in the cat state basis is
\begin{eqnarray}
_z \langle 0 | \Psi \rangle &=& \frac{\Psi_0(\bar z)}{\sqrt{\cosh |z|^2}} = \frac{\Psi(\bar z) + \Psi(-\bar z)}{2 \sqrt{\cosh |z|^2}},\nonumber \\
_z \langle 1 | \Psi \rangle &=& \frac{\Psi_1(\bar z)}{\sqrt{\sinh |z|^2}} = \frac{\Psi(\bar z) - \Psi(-\bar z)}{2 \sqrt{\sinh |z|^2}}.\nonumber
\end{eqnarray}
These wave functions determine even and odd entire analytic functions
\begin{eqnarray}
\Psi_0(z) &=& \frac{1}{2}(\Psi(z) + \Psi(-z)),\\
\Psi_1(z) &=&\frac{1}{2}(\Psi(z) - \Psi(-z)).
\end{eqnarray}
For Glauber coherent state $|\alpha \rangle$, this gives
\begin{eqnarray}
\Psi_0(z) &=& \frac{e^{\alpha z} + e^{-\alpha z}}{2} = \cosh \alpha z,  \label{catFB0} \\
\Psi_1(z) &=& \frac{e^{\alpha z} - e^{-\alpha z}}{2} = \sinh \alpha z. \label{catFB1}
\end{eqnarray}
The corresponding complex potential and complex velocity of the vortex flow are
\begin{eqnarray}
f_0(z) &=& \frac{i\Gamma}{2\pi} \ln \cosh \alpha z,\hskip1cm \bar v_0(z) =\frac{i\Gamma}{2\pi} \alpha \tanh \alpha z,\\
f_1(z) &=& \frac{i\Gamma}{2\pi} \ln \sinh \alpha z,\hskip1cm  \bar v_1(z) =\frac{i\Gamma}{2\pi} \alpha \coth \alpha z.
\end{eqnarray}

By using infinite product representation of hyperbolic functions
\begin{eqnarray}
\sinh \alpha z &=& \alpha z \prod^\infty_{n=1} \left( 1 + \frac{\alpha^2 z^2}{(\pi n)^2}\right),\\
\cosh \alpha z &=& \prod^\infty_{n=1} \left( 1 + \frac{\alpha^2 z^2}{(\pi n - \frac{\pi}{2})^2}\right),
\end{eqnarray}
 and omitting irrelevant constants we have following proposition.
 
 {\bf Proposition 7.}
 The complex potentials of vortex flow in cat states are
\begin{eqnarray}
f_0(z) &=&\frac{i\Gamma}{2\pi} \sum^{\infty}_{n=-\infty} \ln \left( z + i \frac{\pi}{\alpha} (n + \frac{1}{2})   \right),\\
f_1(z) &=&\frac{i\Gamma}{2\pi} \sum^\infty_{n=-\infty} \ln \left(z - i \frac{\pi}{\alpha} n \right),
\end{eqnarray}
with corresponding velocities
\begin{eqnarray}
\bar v_0(z) &=& \frac{i\Gamma}{2\pi} \sum^\infty_{n=-\infty} \frac{1}{z + i\frac{\pi}{\alpha}(n + \frac{1}{2})}, \\
\bar v_1(z) &=& \frac{i\Gamma}{2\pi} \sum^\infty_{n=-\infty} \frac{1}{z - i\frac{\pi}{\alpha}n}.
\end{eqnarray}

These formulas show that both flows are periodic
$$
f_0\left(z+ i \frac{\pi}{\alpha}\right) = f_0(z), f_1\left(z+ i \frac{\pi}{\alpha}\right) = f_1(z),
$$
and describe one dimensional lattice with equal strength vortices, distributed with period $\frac{\pi}{|\alpha|}$ along the line from origin, inclined by angle
$\beta = \frac{\pi}{2} - \arg \alpha$.

{\bf Proposition 8.}
The source/sink flows corresponding to cat states are
\begin{eqnarray}
f_0(z) &=&\frac{N}{2\pi} \sum^{\infty}_{n=-\infty} \ln \left( z + i \frac{\pi}{\alpha} (n + \frac{1}{2})   \right),\\
f_1(z) &=&\frac{N}{2\pi} \sum^\infty_{n=-\infty} \ln \left(z + i \frac{\pi}{\alpha} n \right).
\end{eqnarray}

The flows are also periodic  and like in the vortex representation, describe one dimensional lattice of sources/sinks. But in contrast to the vortex case,
these flows allow interpretation in terms of the oblique strip theorem (\ref{obstriptheorem}),  as just one source/sink in bounded domain in form of the oblique strip, which is determined by complex parameter $\alpha$ of the coherent state.

{\bf Theorem 6.}
The source/sink flow corresponding to the cat state $|1\rangle_\alpha$ represents one point source/sink at origin in the oblique strip domain  $(-\infty < x < \infty)$,
\begin{equation}S_{\alpha}: \{  z = x + i y;  - x \tan \arg \alpha - \frac{\pi}{2 |\alpha|\cos \arg\alpha}   \le  y \le -x \tan \arg \alpha + \frac{\pi}{2 |\alpha|\cos \arg\alpha}\}, \label{ostrip} \end{equation}
the symmetric strip on interval $[-\frac{\pi}{2|\alpha|}, \frac{\pi}{2|\alpha|}]$, with width $h =\frac{\pi}{|\alpha|}$ , and inclined by angle $\beta = - \arg \alpha$.

{\bf Theorem 7.}
The source/sink flow corresponding to the cat state $|0\rangle_\alpha$ represents one point source/sink in the shifted oblique strip domain (\ref{ostrip}), with interval $[0, \frac{\pi}{|\alpha|}]$,
which implies translation $z \rightarrow z + i \frac{\pi}{2 |\alpha|}$.

 The cat states represent orthogonal superposition of coherent states and have many applications from quantum optics to quantum information theory. The above theorems provide interpretation of
these states as one point source/sink for the hydrodynamic flow in the strip domain.

\subsection{Qutrit Coherent States Flow }
The qutrit coherent state superposition \cite{P4}, \cite{PashaevKocak}, \cite{KocakPashaev} in Fock-Bargmann representation is given by three complex functions
\begin{eqnarray}
\Psi_0(z) &=& \frac{1}{3}\left( e^{\alpha z} + e^{\alpha z e^{i\frac{2\pi}{3}}} +   e^{\alpha z e^{-i\frac{2\pi}{3}}} \right),\\
\Psi_1(z) &=& \frac{1}{3}\left( e^{\alpha z} + e^{-i\frac{2\pi}{3}} e^{\alpha z e^{i\frac{2\pi}{3}}} +  e^{i\frac{2\pi}{3}} e^{\alpha z e^{-i\frac{2\pi}{3}}} \right),\\
\Psi_2(z) &=& \frac{1}{3}\left( e^{\alpha z} + e^{i\frac{2\pi}{3}} e^{\alpha z e^{i\frac{2\pi}{3}}} +  e^{-i\frac{2\pi}{3}} e^{\alpha z e^{-i\frac{2\pi}{3}}} \right),
\end{eqnarray}
satisfying
\begin{equation}
\frac{d}{dz} \Psi_0(z) = \alpha \,\Psi_2(z),\hskip0.5cm \frac{d}{dz} \Psi_2(z) = \alpha \,\Psi_1(z),\hskip0.5cm \frac{d}{dz} \Psi_1(z) = \alpha \,\Psi_0(z).
\end{equation} 
This implies corresponding velocity vector flows as ratio of the wave functions
\begin{eqnarray}
\bar v_0(z) &=& \frac{i\Gamma}{2\pi} \frac{\Psi'_0}{\Psi_0} = \frac{i\Gamma}{2\pi} \alpha\frac{\Psi_2}{\Psi_0}, \\
\bar v_2(z) &=& \frac{i\Gamma}{2\pi} \frac{\Psi'_2}{\Psi_2} = \frac{i\Gamma}{2\pi} \alpha\frac{\Psi_1}{\Psi_2}, \\
\bar v_1(z) &=& \frac{i\Gamma}{2\pi} \frac{\Psi'_1}{\Psi_1} = \frac{i\Gamma}{2\pi} \alpha\frac{\Psi_0}{\Psi_1}.
\end{eqnarray} 
The wave functions transform under rotation on angle $\frac{2\pi}{3}$ as
\begin{equation}
\Psi_0(e^{i\frac{2\pi}{3}} z) = \Psi_0(z),\hskip0.5cm \Psi_1(e^{i\frac{2\pi}{3}} z) = e^{i\frac{2\pi}{3}}\Psi_1(z),\hskip0.5cm \Psi_2(e^{i\frac{2\pi}{3}} z) = e^{-i\frac{2\pi}{3}}\Psi_2(z).
\end{equation}
This means that corresponding complex potentials 
\begin{equation}
f_s(z) = \frac{i\Gamma}{2\pi} \ln \Psi_s(z),\,\,\,s=0,1,2,
\end{equation}
are transformed as
\begin{eqnarray}
f_0 (e^{i\frac{2\pi}{3}} z) &=& f_0(z),\\
f_1 (e^{i\frac{2\pi}{3}} z) &=& f_1(z) - \frac{\Gamma}{3}, \\
f_2 (e^{i\frac{2\pi}{3}} z) &=& f_2(z) + \frac{\Gamma}{3},
\end{eqnarray}
and the flow is invariant under these rotations
\begin{eqnarray}
\bar v_s( e^{i\frac{2\pi}{3}} z) = \frac{d}{d z} f_s(e^{i\frac{2\pi}{3}}  z) = \frac{i\Gamma}{2\pi} \frac{d^2}{d z^2} \ln \Psi_s(z) = \bar v_s(z),\,\,\,s =0,1,2
\end{eqnarray}
\section{q-Deformed Coherent States}
The pair of q-bosonic operators $a_q$ and $a_q^+$ satisfy the commutation relation
$$
a_q a_q^+ - q \,a_q^+ a_q =1,
$$
and determines the q-Fock states \cite{arik},
\begin{equation}
|n \rangle_q = \frac{(a_q^+)^n}{\sqrt{[n]_q!}} |0 \rangle_q, \hskip1cm a_q |0\rangle_q =0,
\end{equation}
where non-symmetric q-number is $[n]_q = 1 + q + q^2 + ... q^{n-1} = \frac{q^n -1}{q-1}$.

{\bf Definition 5.}
The q-deformed  coherent states (not normalized) are vectors in q-Fock space
 \begin{equation}
|\alpha \rangle_q = \sum^\infty_{n=0} \frac{\alpha^n}{\sqrt{[n]_q!}} |n\rangle_q , \label{coherent}
\end{equation}
where $\alpha \in C$ is an arbitrary complex number.

{\bf Proposition 9.}
The wave function of the q-deformed coherent state $|\alpha \rangle_q$ in q-Fock-Bargmann representation is
\begin{equation}
_q\langle z | \alpha \rangle_q = \Psi_q(\bar z) \,\,\,    \rightarrow   \,\,\,     \Psi_q(z) = e_q^{\alpha z},
\end{equation}
where $e^z_q$ is the Jackson exponential function \cite{Kac}.

For $|q| > 1$ this function is entire and can be represented by infinite product
\begin{equation}
e_q^{\alpha z} = \prod^\infty_{k=0} \left( 1 + \frac{\alpha z}{q^k} \left(1 - \frac{1}{q}\right) \right) .
\end{equation}
For $|q| < 1$ the function is analytic in the disk $|z| < \frac{1}{\sqrt{1 - q}}$, and  meromorphic, with infinite number of poles
\begin{equation}
e_q^{\alpha z} = \prod^\infty_{k=0} \frac{1}{( 1 - q^k (1-q) \alpha z)  }   .
\end{equation}

{\bf Theorem 7.}
The vortex flow for q-deformed coherent states, for $|q| > 1$
\begin{equation}
f_q(z) = \frac{i \Gamma}{2 \pi} \sum^\infty_{k=0} \ln \left( z + \frac{q^k}{\alpha (q-1)} \right)
\end{equation}
represents an infinite set of point vortices with positions at geometric progression $z_k = - \frac{q^k}{\alpha (q-1)}$,
and for $|q| < 1$
\begin{equation}
f_q(z) = -\frac{i \Gamma}{2 \pi} \sum^\infty_{k=0} \ln \left( z - \frac{1}{\alpha (1-q) q^k} \right)
\end{equation}
by the anti-vortices at positions $z_k = \frac{1}{\alpha (1-q) q^k}$.

An infinite set of point vortex images, located along the line in geometric progression appears in method of images for hydrodynamic flow applied to two concentric circles, the problem discussed seems
the first time in lectures of Poincare \cite{Poincare}.
These images were calculated by the two circle theorem \cite{P1} and represented in terms of q-calculus by Jackson's exponential function in\cite{PY}. These results were motivation for us to search the relation between
hydrodynamic flow in bounded domain and quantum states, representable by complex analytic functions with quantum group symmetry.

\section{Conclusions}
The Madelung fluid representation provides a basis for several interpretations of quantum mechanics and becomes fundamental tool in description of quantum fluids and nonlinear evolution equations. It
has been applied as computational tool for solving non-stationary quantum problems, when instead of following quantum trajectories of individual particles, both trajectories and hydrodynamic fields are computed by fluid dynamics computational techniques. In present paper a new approach is proposed, to relate quantum theory with fluid dynamics. The key point of this approach is Fock-Bargmann representation of the wave function by complex analytic function,
which provides a bridge between these two subjects.

{\bf Acknowledgments}
This work was supported by Izmir Institute of Technology, BAP project 2022IYTE-1-0002.

%
%

\end{document}